\documentclass[12pt,osajnl2,preprint,showpacs,superscriptaddress]{revtex4}  
\usepackage[usenames]{color}
\usepackage{amsmath}
\begin{document}

\title{Atom-photon interactions in a system of coupled cavities}

\author{A. El Amili}
\altaffiliation{Present address: Institut de Physique de Rennes, Universit\'e de Rennes 1-CNRS UMR 6251, Campus de Beaulieu, 35042 Rennes Cedex, France.}
\author{S. Gleyzes}
\altaffiliation{Present address: Laboratoire Kastler Brossel, ENS, UMPC-Paris 6, CNRS, 24 rue Lhomond, 75005 Paris, France.}
\author{C. I. Westbrook}
\email{christoph.westbrook@institutoptique.fr}
\affiliation{Laboratoire Charles Fabry, Institut d'Optique, CNRS et Universit\'e Paris-Sud, Campus Polytechnique, 91127 Palaiseau Cedex, France.}

\begin{abstract}
We give a theoretical treatment of single atom detection in an compound, optical micro cavity. The cavity consists of a single mode semiconductor waveguide with a gap to allow atoms to interact with the optical field in the cavity. Optical losses, both in the semiconductor and induced by the gap are considered and we give an estimate of the cavity finesse. We also compute the cooperativity parameter and show how it depends on the gap width and cavity length. Maximization of the cooperativity does not always correspond to maximization of the coupling.
\end{abstract}

\ocis{020.5580,130.3120,230.3120,230.4555,270.5570}
\maketitle

\section{Introduction}

Recent years have seen significant developments in atom optics because of miniaturization, which has allowed workers to trap and manipulate atomic ensembles near a substrate via magnetic fields generated by currents in microfabricated wires \cite{Fortagh,Reichel1,Denschlag}. These advances have permitted generation of Bose-Einstein condensates on atom chips, 
and production of coherent sources of atoms \cite{Hansel,ott,Schneider}. These scalable micro structures promise various applications on quantum information processing and micro-interferometry on a chip \cite{Enk,WangMichelson,Schumm}. In both cases, an important step is to implement a local atom detector able to detect low atom numbers - ideally single atoms. Several detection possibilities have been, theoretically and experimentally, demonstrated using the interaction of atoms with resonant light \cite{Horak,Poldy,Steinmetz,Trupke,Colombe,Kohnen}. For detection in an optical cavity, an important figure of merit is the ``cooperativity". The cooperativity $C = g^2/\kappa\gamma$ describes the competition between the atom-field coupling $g$ and the damping rates given by the half linewidth of the cavity and the atomic transitions ($\kappa$, $\gamma$) respectively \cite{Kimble, Horak,Poldy}. In other words, $C^{-1}$ is the number of atoms needed to significantly disturb the cavity.
In a simple Fabry-P\'erot cavity, the cooperativity $C\propto \mathcal{F}/\mathcal{A}$ depends mainly on the cavity parameters such as the finesse $\mathcal{F}$ and the intra-cavity beam cross section $\mathcal{A}$. Even for a low cavity finesse, it is possible to reach useful values of the cooperativity if the intra-cavity cross section is small. In a system of several coupled cavities, as we will analyse here, the cooperativity is not a simple function of the  finesse and beam cross section. 
This is because the electric field in different parts of the cavity can be very different and can depend on other parameters such as the lengths of the various sub-cavities. We therefore undertake in this paper to analyse a specific coupled cavity system with a view towards optimizing the cooperativity. We will show that the optimum operating conditions are not always easy to guess.

In a previous paper~\cite{Gleyzes}, we described a cavity consisting of an AlGaAs waveguide with mirrors at both ends. A cross section of the waveguide is shown in Fig.~\ref{fig-dessin} (a). To use this device as an atom detector, we propose to open a small gap in the waveguide. Atoms within the gap can be detected or manipulated via their interaction with the intracavity field. Because of the considerable effective refractive index of AlGaAs ($n \approx 3.1$), the gap-vacuum interface has a  reflectivity of $27~\%$ and the gap thus forms a low finesse cavity within the larger waveguide cavity. The various length scales of the entire system are shown in Fig.~\ref{fig-dessin} (b). In order to avoid excessive atom-surface interactions, it is necessary that the gap width $d$ be at least $2~\mu$m~\cite{Gleyzes}. On the other hand, the minimization of diffraction losses within the gap dictates that it be as narrow as possible. We have therefore concentrated on analyzing the characteristics of this system for gap sizes of about $2~\mu$m. In our earlier work we gave an estimate of the cooperativity in the regime where the finesse is limited by the waveguide propagation losses. In the present paper we discuss a more systematic investigation. First we study numerically and analytically the transmission and reflection of the gap as a function of its width, taking into account the losses due to diffraction. Knowing the transfer matrix of the gap, we estimate in section \ref{section-finesse} the finesse of a resonator based on such a opened waveguide. In section \ref{section-field}, we compute the amplitude of the field in each part of the resonator in order to calculate the coupling parameter $g$ of an atom to the light field. These calculations allow us to calculate the cooperativity and optimize it as a function of the cavity parameters.

We confirm the approximate results of Ref.~\cite{Gleyzes}.
However, we believe it should be possible to fabricate a waveguide with sufficiently low loss
that the total losses are dominated by the unavoidable loss in the gap. 
In this case, we show that the optimization of the cavity is very different.

\section{Transmission and reflection of the gap}
\label{section-rgtg}

\begin{figure}[ht]
 \centering
 \includegraphics[width=0.8\linewidth]{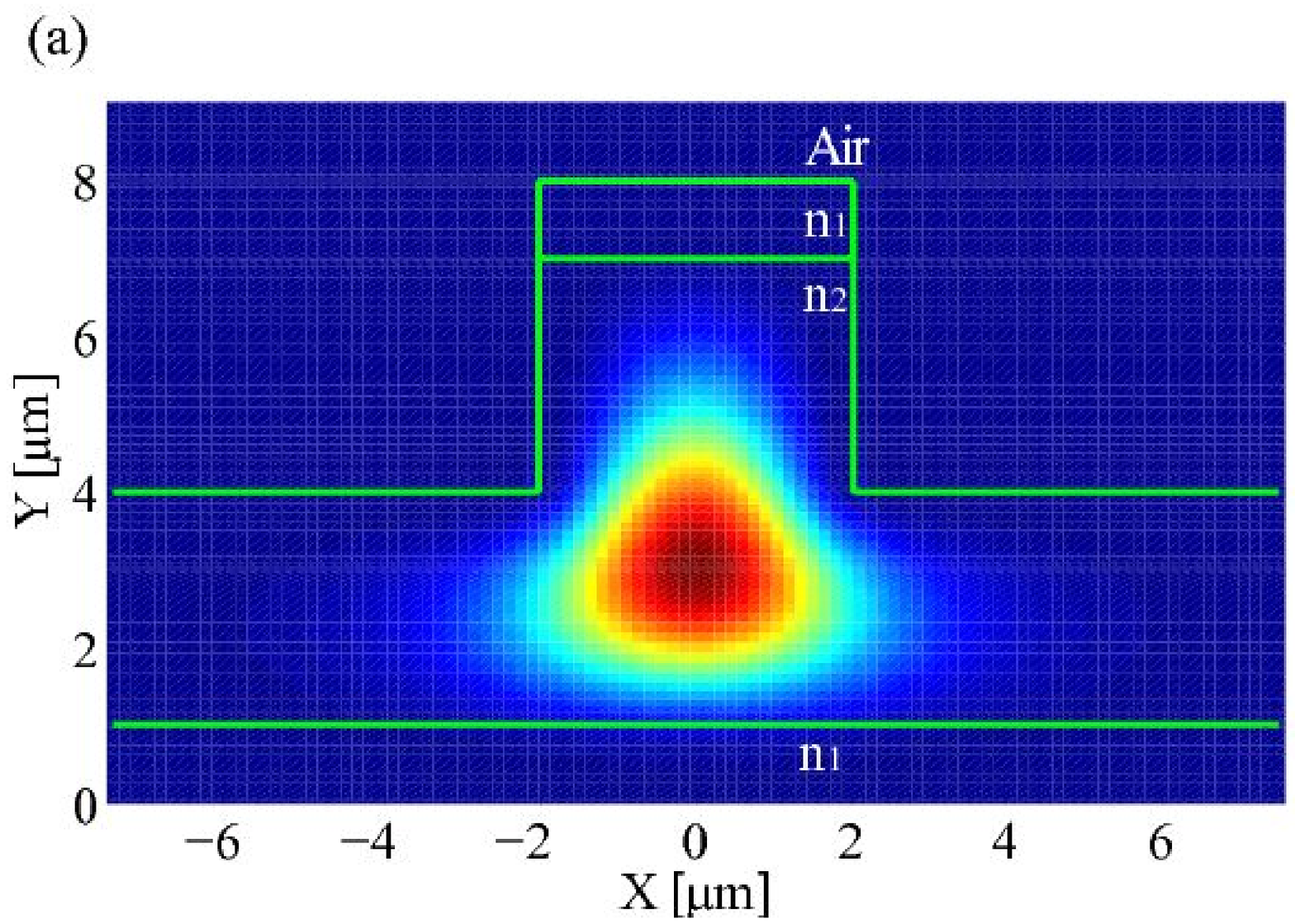}
 \centering
 \includegraphics[width=1\linewidth]{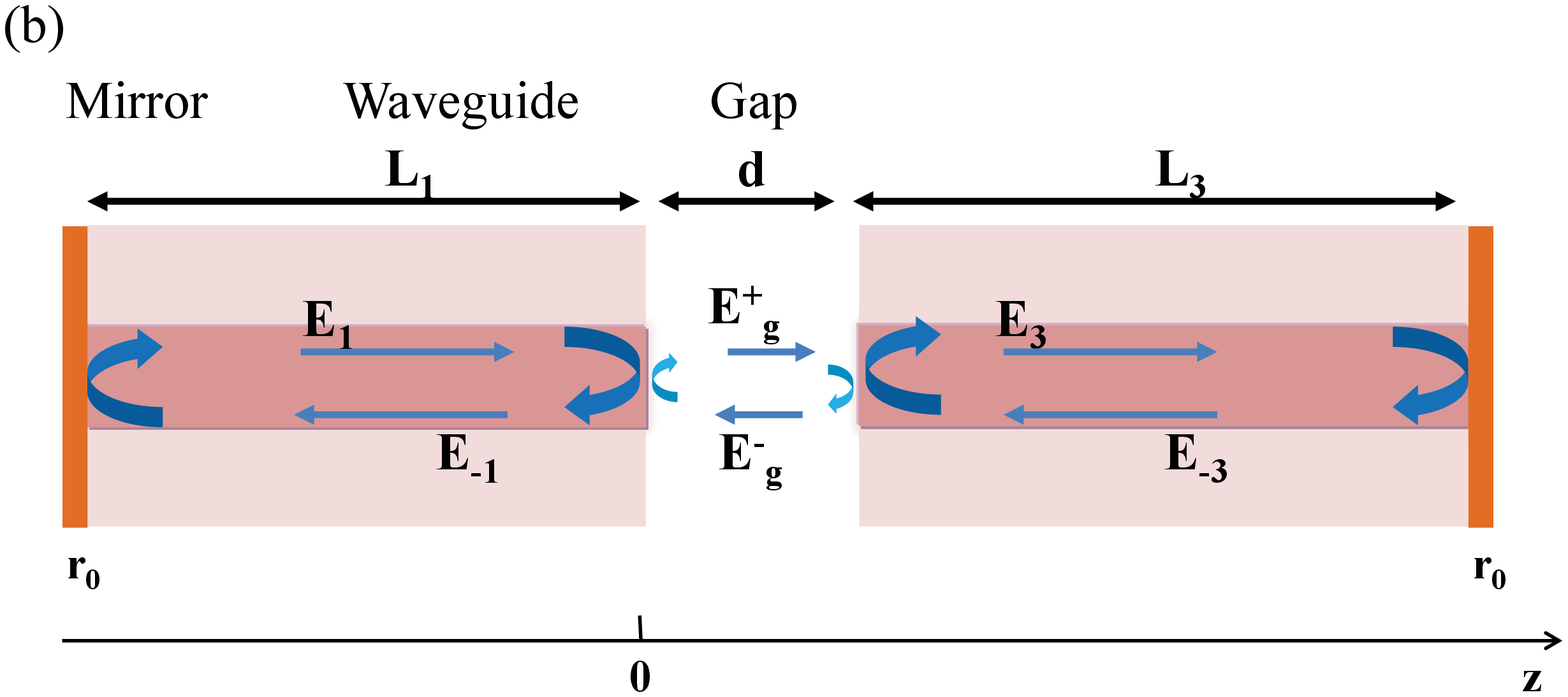}
  \caption{
 (a) Transverse geometry of the waveguide. The waveguide is a rectangular ridge of AlGaAs, $4~\mu$m wide. 
The false color image shows the result of a numerical calculation of the amplitude of fundamental mode of the waveguide operating at $\lambda = $780 nm. The mode area is calculated numerically and we find $\mathcal{A} =$ 9.9 $\mu$m$^2$.
 (b) Top view of the gapped waveguide. Atoms are to be guided into this gap to interact with the intracavity field. 
  The gap constitutes a low finesse cavity within a larger waveguide cavity which is closed by reflecting surfaces at the both ends of the waveguide.
}
  \label{fig-dessin}
\end{figure}

We begin by giving some more details of the calculations of the losses induced by the gap which allowed us to estimate the cooperativity in our previous paper \cite{Gleyzes}. To perform our analysis, a key issue is the spatial variation of the eigenmode of the waveguide as it propagates in free-space. Because of the complex geometry of the waveguide (Fig. \ref{fig-dessin} (a)), the spatial mode profile must be calculated numerically. We consider a field that propagates inside the waveguide for $z < 0$, and that reaches at $z = 0$ a waveguide/vacuum interface (Fig. \ref{fig-dessin} (b)). Using a 3 dimensional fully-vectorial aperiodic-Fourier modal method \cite{Hugonin}, we compute the complex amplitude $\mathcal{E}_{0}(x,y,z)$ of the field that propagates in free-space in the half-space corresponding to $z > 0$. Numerically, we perform the simulation from $z = 0$ to $z = 35~\mu$m. We define the ``axis of the field" as the coordinates $(x_0, y_0)$ that correspond to the maximum of $|\mathcal{E}_{0}(x,y,0)|$ and set the origin of the coordinates so that $x_0 = y_0 = 0$. We define $E_0 = \mathcal{E}_{0}(0,0,0)$, and then rewrite $\mathcal{E}_{0}(x,y,z) =  E_{0} f(x,y,z)$, where $f$ is the mode profile during it's propagation along the $z$-axis. Due to the continuity of the electric field, $f(x,y,0)$ corresponds to the profile of the eigenmode inside the waveguide. We can therefore compute the overlap between the propagated field and the field inside the waveguide, which is expressed as:
\begin{equation}\label{eqn-overlap}
    Q(z) = \frac{\int f^*(x,y,0) f(x,y,z) \mathrm d x  \mathrm d y}{\sqrt{\int |f(x,y,0)|^2  \mathrm d x  \mathrm d y\int |f(x,y,z)|^2 \mathrm d x  \mathrm d y}}.
\end{equation}
For $z>0$, the overlap function is $|Q(z)|<1$, which expresses the fact that because of the diffraction during the propagation in free-space, the field cannot be totally recoupled into the waveguide on the other side of the gap. The knowledge of $f(x,y,z)$ also allows us to compute the phase and amplitude of the electric field on the axis of the waveguide as the field propagates:
\begin{equation}\label{eqn-cpc}
    f(0,0,z) = |f(0,0,z)| \exp i(kz+\varphi_{0}(z)),
\end{equation}
where $\varphi_0$ is the correction to the plane wave phase and varies slowly with $z$ on a scale $\lambda$.
Once we know the free-space propagation of the eigenmode, we can calculate the transmission and reflection of a gap of width $d$. We consider the gap as two plane interfaces separated by a distance $d$. Each interface has a transmission and a reflection given by the Fresnel coefficient of reflection and transmission between a material of refractive index $n=3.15$ and the vacuum $n_{0}=1$. On the left interface, the light coupled into the gap has the spatial profile of the eigenmode of the waveguide. The mode inside the cavity is considered as the superposition of $\mathcal{E}^+_p$ and $\mathcal{E}^-_p$,
the fields corresponding to the $2p^{\mathrm{th}}$ (propagating from left to right) and $2p+1^{\mathrm{th}}$ (propagating from right to left) reflections at the gap interfaces. The amplitude $\mathcal{E}^+_p$ ($\mathcal{E}^-_p$) can be expressed as a function of the incident field amplitude $E_i$ as $E^+_p = tr^{2p}E_i$ and $E^-_p = tr^{2p+1}E_i$. The Fresnel coefficients $t$ and $r$ correspond respectively to the transmission coefficient for a wave travelling from the waveguide into the vacuum, and the reflection coefficient of the vacuum/waveguide interface. The spatial profile of $\mathcal{E}^+_p$ and $\mathcal{E}^-_p$ is related to $f(x,y,z)$ by:
\begin{eqnarray}
\mathcal{E}^+_p(x,y,z) & =&  E^+_p f(x,y,2pd+z)\\
\mathcal{E}^-_p(x,y,z) & =&  E^-_p f(x,y,(2p+2)d-z).
\end{eqnarray}
\begin{figure}[ht]
\centering
\includegraphics[width=0.8\linewidth]{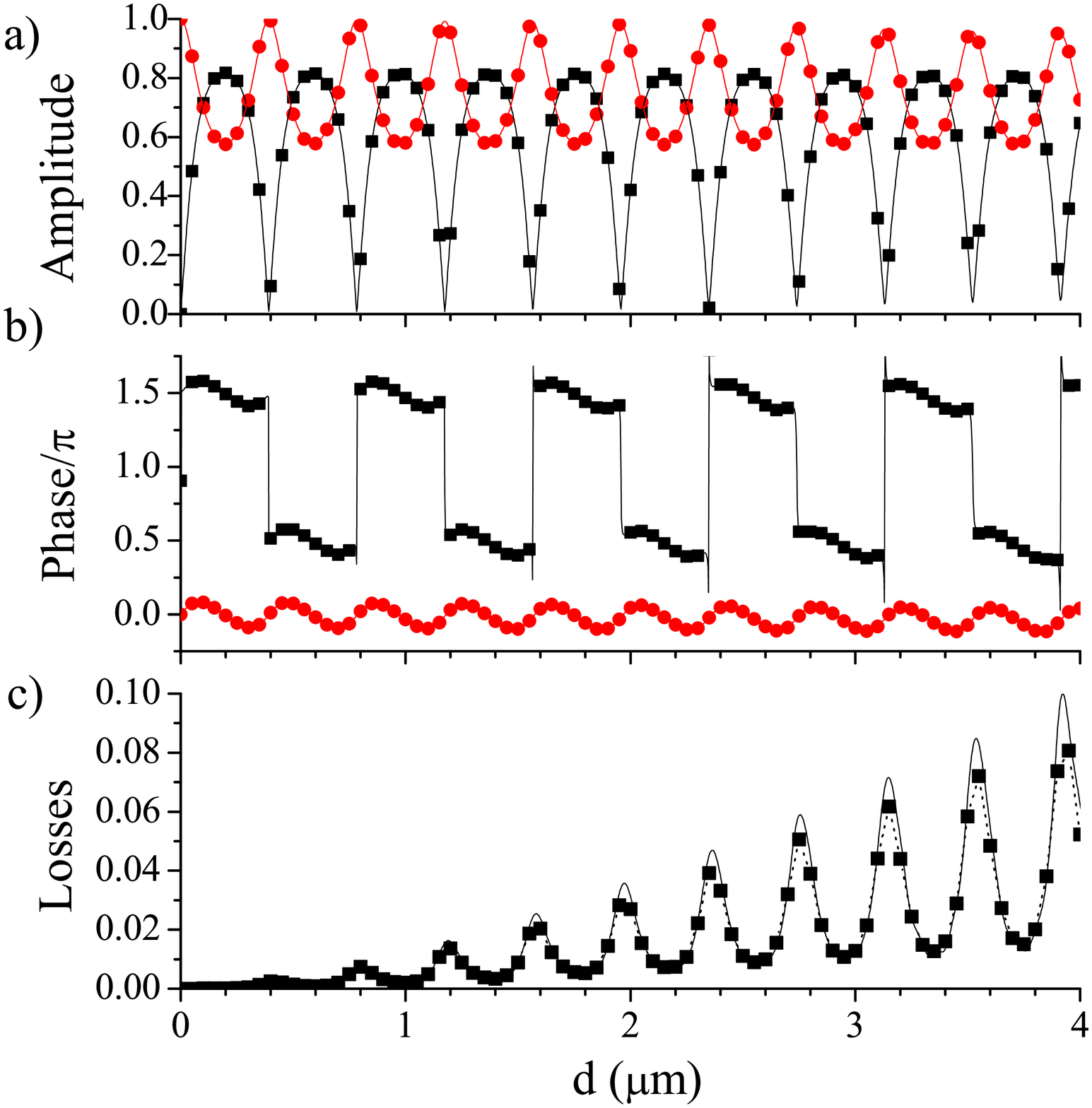}\\
  \caption{(a) Amplitude of the transmission and reflection of the gap as a function of the gap width. (b) Phase of the transmission and reflection coefficients (the propagating term $k_0d$ has been subtracted). The solid lines correspond to the analytical model ($\tilde{r}_g$, $\tilde{t}_g$), the square and circles correspond to the result of the full numerical simulation ($\tilde{r}_{g,sim}$ and $\tilde{t}_{g,sim}$ respectively, see text). (c) Losses induced by the gap. The solid line correspond to $1-|\tilde{t}_g|^2-|\tilde{r}_g|^2$, the dotted line + squares correspond to $1-|\tilde{t}_{g,sim}|^2-|\tilde{r}_{g,sim}|^2$. The losses are maximal for a gap width equal to an integer number of half wavelengths.
  }\label{fig-rgtgt}
\end{figure}
A fraction $t^{'}$, where $t^{'}$ is the transmission coefficient for a wave travelling from the gap into the waveguide, of each component $\mathcal{E}^+_p$ is transmitted through the right interface, but since the light diffracts during its propagation in the free-space of the gap, only a fraction $Q^+_p = Q(2pd +d)$ of the transmitted field is coupled back into the eigenmode of the right waveguide. The amplitude $E_t$ of field transmitted through the gap is therefore:
\begin{equation}
E_t = t^{'} \sum_p Q^+_p t r^{2p} E_i.
\end{equation}
In a similar way, one can write the amplitude of the field reflected by the gap as the sum of the amplitude of the light directly reflected by the waveguide/vacuum interface $-r E_i$ and the mode-matched part of $\mathcal{E}^-_p$ that is transmitted back into the left part of the waveguide:
\begin{equation}
E_r = -r E_i + t^{'} \sum_p Q^-_p t r^{2p+1} E_i,
\end{equation}
where $Q^-_p = Q(2pd+2d)$. 

Considering the gap as a single entity, its complex reflection and transmission coefficients for the amplitude are $\tilde{r}_g =  E_r/E_i$ and $\tilde{t}_g = E_t/ E_i$, and characterize completely the gap. The phase and amplitude of $\tilde{r}_g$ and $\tilde{t}_g$ are shown in Fig. \ref{fig-rgtgt}. To confirm our results, we can also use the simulation software to directly simulate the propagation through a wave\-guide - vacuum - wave\-guide stack. We denote by $\tilde{r}_{g,sim}$ and $\tilde{t}_{g,sim}$ the reflection and transmission coefficients for the amplitude deduced from the numerical simulations. Figures \ref{fig-rgtgt} (a) and  \ref{fig-rgtgt} (b) compare the result of the analytical model to the numerical simulations. We see that the amplitude and phase match very well. We also compared the losses, defined as $1-|\tilde{t}_g|^2-|\tilde{r}_g|^2$, calculated using the two methods and they agree within $20~\%$.

\section{Resonator losses induced by the gap}
\label{section-finesse}

The losses calculated above only correspond to the losses for a single pass inside the cavity. The losses induced by the gap in a cavity will depend on the interferences between the two propagating fields that create the standing wave inside the resonator. Therefore, to calculate the maximum finesse of a cavity based on a gapped waveguide, we need to estimate the losses induced by the gap in the presence of the counter-propagating field reflected by a perfect mirror that terminates the cavity (reflectivity $r_{0} = -1$) (Fig. \ref{fig-dessin} (b)). We denote by $E_1$ ($E_3$) the amplitude just before (after) the gap of the field propagating in the left (right) waveguide from the left to the right, and $E_{-1}$ ($E_{-3}$) the amplitude just before (after) the gap of the field in the left (right) waveguide propagating from right to left. Following the notation of section \ref{section-rgtg}, one can write:
\begin{eqnarray}
  E_{3}  &=& \tilde{t}_g(d) E_{1}+\tilde{r}_g(d) E_{-3}   \label{eqn-aaaa1} \\
  E_{-1} &=& \tilde{t}_g(d) E_{-3}+\tilde{r}_g(d) E_{1}   \label{eqn-aaaa2} \\
  E_{-3} &=& r_0\exp(2inkL_3)E_{3},
  \label{eqn-aaaa3}
\end{eqnarray}
where $n$ is the effective refractive index of the eigenmode of the waveguide and $k=2\pi/\lambda$ the wavenumber of the light in vacuum. Solving (\ref{eqn-aaaa1}$-$\ref{eqn-aaaa3}), we can express $E_{-1}$ as a function of $E_1$:
\begin{equation}\label{eqn-rstar}
    E_{-1} = \tilde{r}_{eff}(d,k,L_3) E_1,
\end{equation}
where $\tilde{r}_{eff}$ is the effective reflectivity of the ensemble gap + waveguide + end mirror. 
Once we have calculated $\tilde{r}_{eff}$, we can calculate the finesse of a resonator with a gap. It is the finesse of the equivalent resonator made of a perfect mirror (reflectivity $r_0=-1$) facing a mirror of reflectivity $\tilde{r}_{eff}$, separated by a distance $L_1$. If we write $\tilde{r}_{eff} = r_{eff} \exp (2i\phi)$, we can express the finesse as:
   \begin{equation}\label{eqn-finesse}
    \mathcal{F}_{d} =\frac{\pi\sqrt{r_{eff}}}{1-r_{eff}} \frac{1}{n(L_1+L_3)} \left ( nL_1+\frac{\partial \phi}{\partial k}\right).\end{equation}

In the following, we assume that $L_3\gg d$. Since we are looking for a resonance of the cavity, we consider the response $r_{eff}$ for wavenumber $k + \delta k$ such that $\delta k\lesssim 1/L_3\ll 1/d $. Therefore we will consider that $\tilde{r}_{g}(k + \delta k)\approx \tilde{r}_g(k)$ and $\tilde{t}_{g}(k + \delta k)\approx \tilde{t}_g(k)$ and that the only $k$-dependence of $\phi$ comes from the $\exp(2inkL_3)$ of Eq. \ref{eqn-aaaa3}.
\begin{figure}[ht]
 \includegraphics[scale=.7]{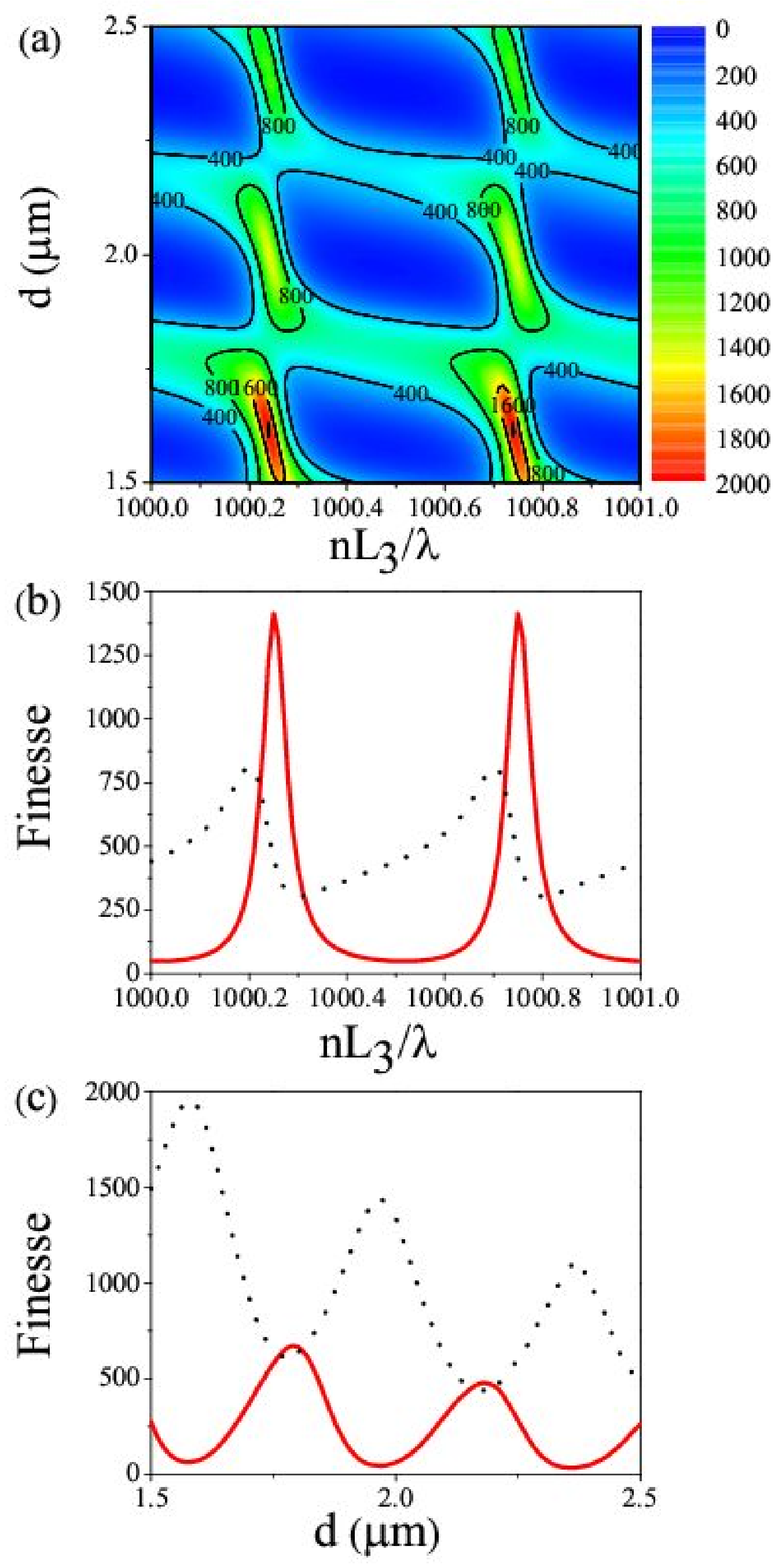}
     \caption{(a) Cavity finesse as function the gap width $d$ and waveguide length $L_3$.
     (b) Finesse as a function of $L_3$ for $d =$ 1.95 $\mu$m (solid red line), and $d \approx$ 2.15 $\mu$m  (dotted black line).
     (c) Finesse as a function of $d$ for $nL_{3}/\lambda =$ 1000.5 (solid red line), and $nL_{3}/\lambda =$ 1000.25 (dotted black line).}
     \label{finesse}
\end{figure}
Knowing the effective reflectivity $\tilde{r}_{eff}$, we are able to compute the finesse of such a compound micro cavity. The results are shown in Fig. \ref{finesse} for different values of $d$, $L_1$ and $L_3$. We varied $d$ between 1.5 and 2.5 $\mu$m, and $L_3$ between $1000\lambda/n$ and $1001\lambda/n$. Once $d$ and $L_3$ are fixed, the distance $L_1$ is no longer a free parameter since it has to be chosen to verify the resonance condition of the cavity. However, we can always choose $L_1$ within $\lambda/2n$ of $1000\lambda/n$. Therefore, in the calculations, we assume that $L_{1}\approx1000\lambda/n$, and only plot the finesse as function of $d$ and $L_3$. The choice of $1000\lambda/n$ is arbitrary and simply ensures the two conditions $L_{3}\gg d$ and $L_{1}\gg \lambda/n$, necessary to neglect the dependence of $L_1$ with $L_3$ and $d$ in the calculations. Since $L_3$ only appears in equations (\ref{eqn-aaaa1}$-$\ref{eqn-aaaa3}) as $\exp(2iknL_{3})$, one can check that
   \begin{equation}\label{eqn-finesse-bis}
\mathcal{F}_{d} =\frac{\pi\sqrt{r_{eff}}}{1-r_{eff}} \frac{1}{n(L_1+L_3)} \left ( nL_1+nL_3\frac{\partial \phi}{\partial\left ( nk\delta L_{3}\right)}\right),
   \end{equation}
where $\delta L_{3} = L_{3}$ mod $\lambda/2n$. Equation \ref{eqn-finesse-bis} clearly shows that $\mathcal{F}_{d}$ does not change if we multiply $L_1$ and $L_3$ by the same factor, as long as we keep $\delta L_{3}$ constant.
\newline 

The plots in Fig. \ref{finesse} show that the finesse globally decreases when the gap width $d$ increases. This is consistent with the result of Fig. \ref{fig-rgtgt}, and corresponds again to the fact that the diffraction losses increase with $d$. In addition to the global decrease, we observe oscillations of the finesse as a function of both $d$ and $L_3$. To understand these oscillations, we recall that the losses correspond to fields that are transmitted across the gap but are not coupled into the guided mode on the other side. In the presence of a perfect right-end mirror, the amplitude of the field leaking out at each interface is the sum of a contribution proportional to $E_{1}$ (the single pass losses of the incident field) and a contribution proportional to $E_{-3}$ (the single pass losses of the field reflected by the end mirror). Depending on the relative phase of $E_1$ and $E_{-3}$, the interference between these two contributions can be either constructive or destructive, and one can reduce the losses in the gap and reach a much larger finesse than what could be deduced from the single pass losses.

\section{Atom-field coupling}
 \label{section-field}
 
\begin{figure}[ht]
\centering
   \includegraphics[width=0.8\linewidth]{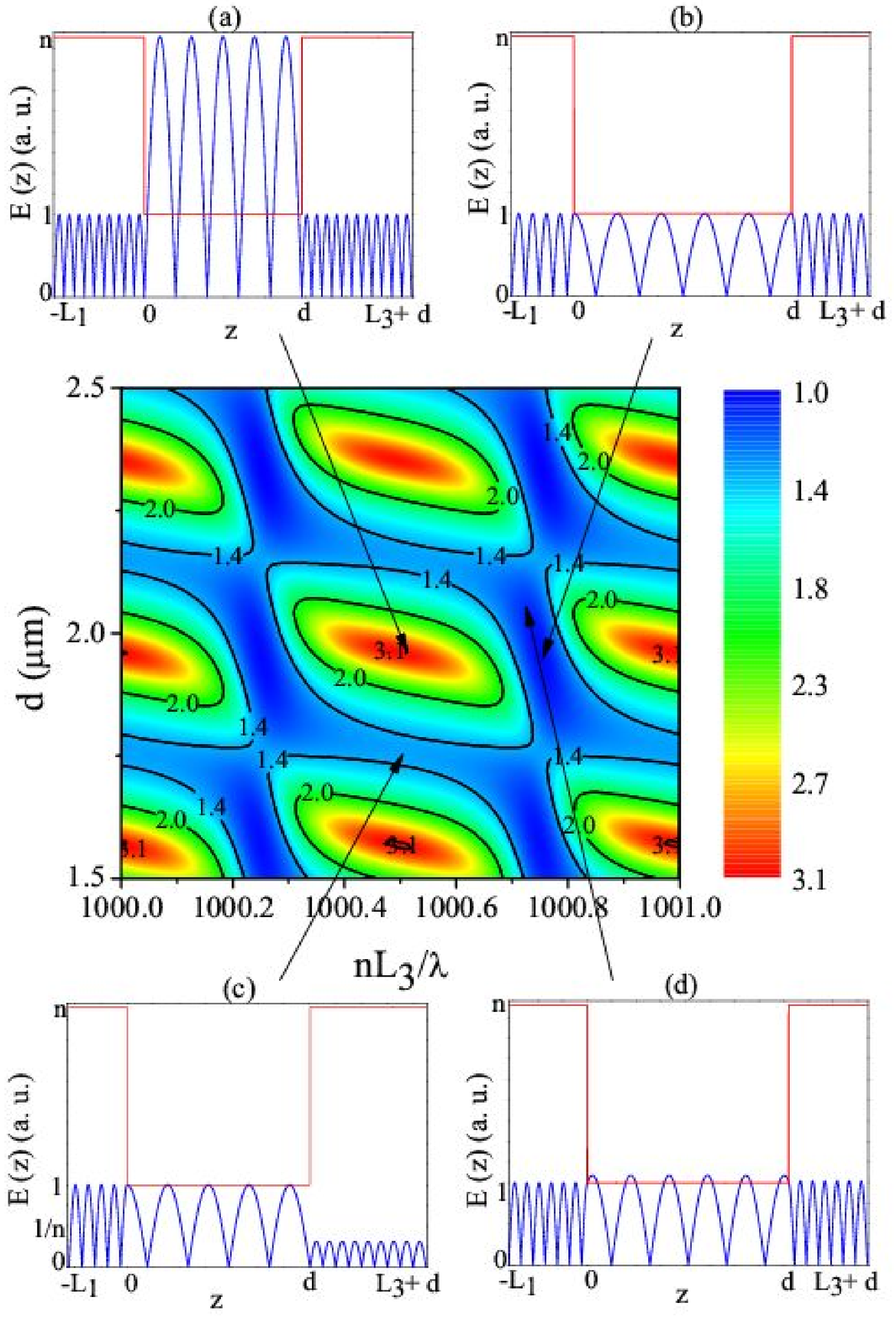}\\
  \caption{Contour plot of the atom-field coupling $\tilde{g}$ as function $d$ and $L_3$. Insets show the electric field profile inside the coupled cavity for special cases of $d$ and $L_3$ with a normalisation $\left|E_{1}\right| + \left|E_{-1}\right| =$ 1. (a) Electric field profile for $d =$ 1.95 $\mu$m and $L_{3}$ equal to an integer number of half wavelengths. (b) Electric field profile for $d =$ 1.95 $\mu$m and $L_{3} =\lambda/4n$ (mod $\lambda/2n$). (c) Electric field profile for $d =$ 1.75 $\mu$m and $L_{3}$ equal to an integer number of half wavelengths. (d) Electric field profile for $d =$ 2.05 $\mu$m and $L_{3} = 0.23\lambda$ (mod $\lambda/2n$) (value that optimizes $\mathcal{C}$ in the lossless case).}
  \label{couplage}
\end{figure}
To estimate the cooperativity of the system, we need to calculate the amplitude of a one-photon field in the center of the gap. 
Following the same argument as in the section \ref{section-finesse}, the field in the gap $E_{g}(z)$ is the sum of the multiple reflections of the light coming from the left waveguide and the multiple reflections coming from the right. Consequently, using equation \ref{eqn-cpc}, one can rewrite $E_{g}(z)$ as the sum of $E^{+}_{g}(z)$ and $E^{-}_{g}(z)$ where
\begin{eqnarray}\label{eqn-fieldgap2}
 \nonumber E^+_{g}(z) &=& \big[E_1\sum_{p\mbox{ }even} t r^{p}  |f_0(pd+z)|\exp(ikpd +i\varphi_0(pd+z)) \\
   \nonumber   & &  +   E_{-3} \sum_{p\mbox{ }odd}  t r^{p} |f_0(pd+z)|\exp(ikpd +i\varphi_0(pd+z))\big]\exp(ikz)   \\
    & = & \tilde E^+_{g}(z)\exp(ikz)\\
  \nonumber  E^-_{g}(z)&=& \big[E_1 \sum_{p\mbox{ }odd}   t r^{p} |f_0(pd+d-z)|\exp(ikpd + i \varphi_0(pd+d-z))  \\
   \nonumber  & &  +  E_{-3} \sum_{p\mbox{ }even} t r^{p} | f_0(pd+d-z)|\exp(ikpd +i\varphi_0(pd+d-z))\big]\exp(-ikz) \\
       & = & \tilde E^-_{g}(z)\exp(-ikz),
\end{eqnarray}
where $|f_{0}(z)| = |f(0,0,z)|$ is the amplitude of the electric field on the axis of the waveguide. Here, $\tilde E^+_{g}$ and $\tilde E^-_{g}$ vary slowly on the scale of $\lambda$. As a result, the field inside the gap is the sum of two terms evolving at $\exp(ikz)$ and $\exp(-ikz)$. The interference of these two terms will not necessarily be constructive at the exact center of the gap $z = d/2$, but there will be an anti-node at a position $z^*$ within $\lambda/2$ of the center where the two amplitudes will add up constructively and
\begin{equation}
    |E_{g}(z^*)| = |\tilde E^+_{g}(z^*)| + |\tilde E^-_{g}(z^*)| \approx |\tilde E^+_{g}\left (\frac d 2\right)| + |\tilde E^-_{g}\left(\frac d 2\right)|.
\end{equation}

Up to this point, the amplitude of the field in the cavity has been set by an arbitrary factor $E_1$. To calculate the single-photon amplitude, one has to write that the total energy stored in the cavity is equal to $\hbar \omega$. Since $d \ll \left\{L_1, L_3\right\}$, we can neglect the energy stored in the gap. The total energy of the cavity is therefore
\begin{equation}
     h\nu \approx 4n^2\epsilon_0 \mathcal{A} (L_1 |E_1|^2 + L_3 |E_3|^2),
\end{equation}
where $\nu= c/\lambda$ is the photon frequency, $\epsilon_0$ is the electric permittivity of vacuum, and we have assumed that $|E_{-1}| \approx |E_1|$ since the losses are small. Finally, the single photon amplitude is given by:
\begin{equation}
    \mathcal{E}= \frac{\sqrt{h\nu}\left( |E^+_{g}| +  |E^-_{g}|\right)}{\sqrt{4n^2\epsilon_0 \mathcal{A} (L_1 |E_1|^2+ L_3 |E_3|^2)}}.
\end{equation}
Assuming that $L_{1}\approx L_{3}\approx L$, we can rewrite the maximum atom field coupling $g$ at the center anti-node as $g = g_{0}\tilde{g}(d,L_{1},L_{3})$ with:

\begin{equation}
g_{0}=\sqrt{\frac{3\pi c\gamma}{k^{2}n^{2}\mathcal{A}L}},
\label{eq-go}
\end{equation}
and
\begin{equation}
\tilde{g}(d,L_{1},L_{3})= \frac{1}{\sqrt{2}}\frac{ |E^+_{g}| + |E^-_{g}|}{\sqrt{|E_1|^2+|E_3|^2}}.
    \label{eq-gtilte}
\end{equation}
Here again, $L_1$ is chosen to ensure that the cavity is resonant at the atomic frequency, and is no longer a free parameter once $d$ and $L_3$ are fixed. We can therefore plot $\tilde{g}(d,L_{1},L_{3}) = \tilde{g}(d,L_{3})$ as function of $d$ and $L_3$ (Fig. \ref{couplage}). The value of $\tilde{g}$ varies between 1 and $n$. This can be very well explained by the inset of the Fig. \ref{couplage} showing the electric field distribution inside the cavity. If $L_1$, $L_3$ and $d$ all correspond to an integer number of half-wavelength of the field, the interferences of the partial reflections at the air-waveguide interfaces enhance the field in the gap by a factor $n$, leading to $\tilde{g} = n$. If $d$ corresponds to a half integer number of half wavelengths in the gap, $\tilde g = \sqrt 2n/\sqrt{n^2+1}$ is constant. If $d$ is an integer number of half wavelengths, but $L_1$ and $L_3$ are equal to $\lambda/4n$ (mod $\lambda/2n$), then the amplitude of the standing wave of the electromagnetic field is constant in the whole cavity and $\tilde g = 1$.
To get a feeling for the magnitude of the coupling, we use the parameters of the above cavity
including the effective mode area $\mathcal{A}=9.9\,\mu\mathrm{m}^2$ (see Ref.~\cite{Gleyzes}).
The formula (\ref{eq-go}) gives $g_0\approx 2 \pi \times 30\, \mathrm{MHz}$ and so the 
maximum value of the coupling is about $2 \pi \times 100\, \mathrm{MHz}$.

\section{Cooperativity of the atom-cavity system}

The figure of merit of the cavity is given by the cooperativity $C=g^2/\kappa\gamma$. The cavity loss rate $\kappa$ is inversely proportional to the finesse:
\begin{equation}
  \kappa=\frac{\pi c}{2n(L_1+L_3)\mathcal{F}}, 
\end{equation}
and the cooperativity is then given by:
\begin{equation}
    C=2\frac{\sigma_{abs}\mathcal{F}}{\pi n\mathcal{A}}\tilde{g}^{2}(d,L_{3}),
    \label{eqn-Cfinal}
\end{equation}
where $\sigma_{abs}= 6\pi/k^2$ is the resonant atomic cross section.

\begin{figure}[ht]
\centering
   \includegraphics[width=0.8\linewidth]{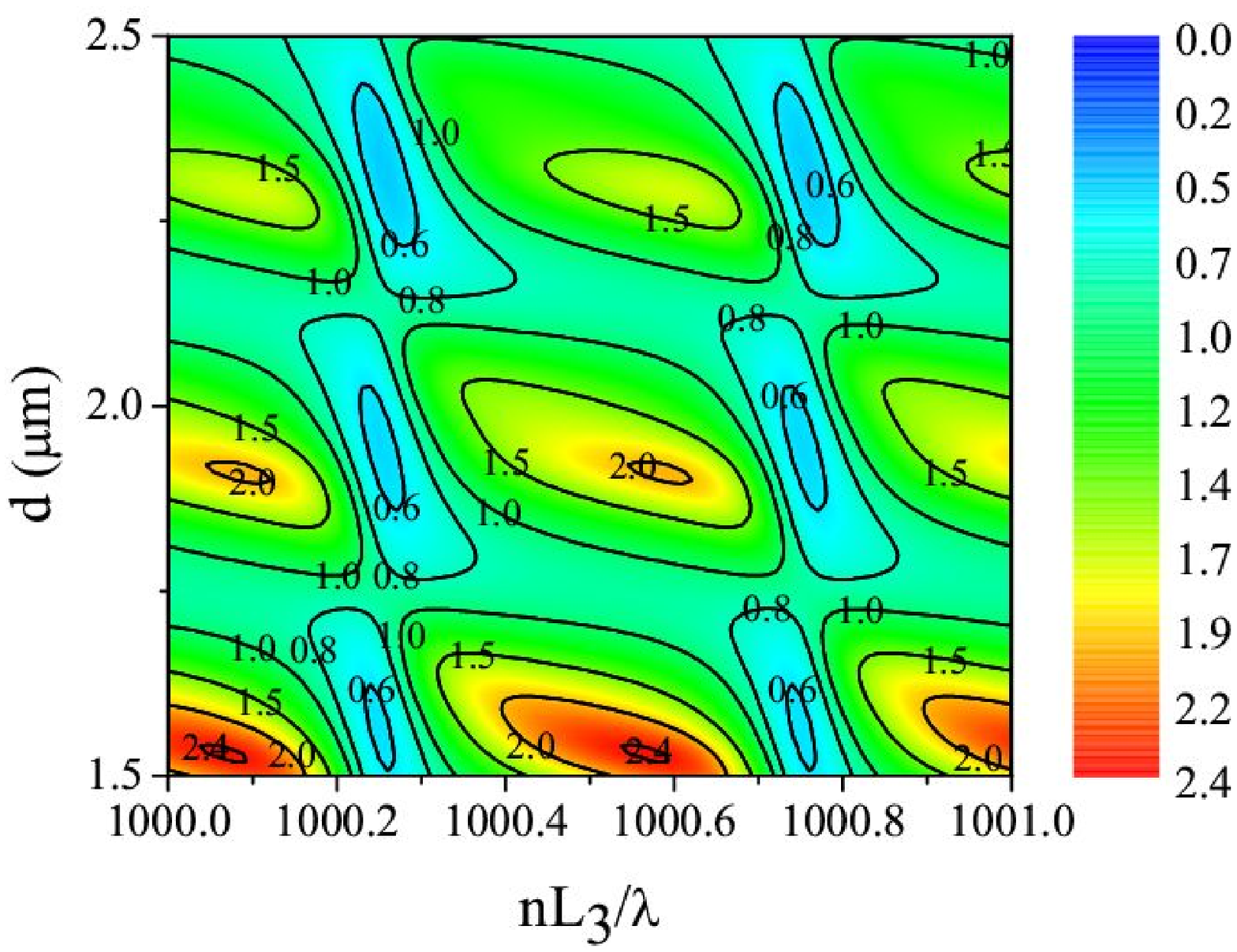}\\
  \caption{Cooperativity of a Rb atom in the coupled cavity system as a function of $d$ and $L_3$, including propagation losses in the waveguides. The attenuation coefficient corresponds to that measured in Ref.~\cite{Gleyzes}. The coupling is maximized for $d =$ 1.95 $\mu$m and $nL_{3} =$ 1000.5$\lambda$.}
  \label{fig-Cfinal-pertes}
\end{figure}

We can see that the cooperativity is proportional to the product $\tilde{g}^{2}\mathcal{F}$. Comparing the results of sections II and III shows that optimizing the cooperativity will be a trade off between maximizing $\mathcal{F}$ and maximizing $\tilde{g}$. The total finesse of the cavity is given by $\mathcal{F}^{-1} = \mathcal{F}^{-1}_{d} + \mathcal{F}^{-1}_{p}$, where $\mathcal{F}_{d}$ corresponds to the finesse induced by the diffraction losses (calculated in section 3), and $\mathcal{F}_{p}$ corresponds to the other intrinsic losses of the cavity (such as the losses due to the absorption or scattering of the light during its propagation in the waveguide part of the cavity). Using our numerical model, we have computed the cooperativity as a function of $d$ and $L_3$ in the two regime $\mathcal{F}_{d} \gg \mathcal{F}_{p}$ and $\mathcal{F}_{d} \ll\mathcal{F}_{p}$. Each time we vary $d$ between 1.5 and 2.5 $\mu$m, and $L_3$ between $1000\lambda/n$ and $1001\lambda/n$ keeping $L_1\approx1000\lambda/n$ in the two different cases.

For the first case, we consider the losses due to  absorption and/ or diffusion in the AlGaAs waveguide as experimentally measured in \cite{Gleyzes}, corresponding to $\mathcal{F}_p \sim 92$. The results are shown Fig. \ref{fig-Cfinal-pertes} and they confirm what was suggested in \cite{Gleyzes}. If the propagation losses are high, the total finesse is limited by $\mathcal{F}_{p}$, and the maximum of the cooperativity is a point in the parameter space very close to the values $d = 1.95~\mu$m and $nL_{3} = 1000\lambda$ which optimize the coupling $\tilde{g}$. Still, the calculation shows that the cooperativity can be increased by $\sim$10~$\%$ by slightly decreasing $d$ and increasing $L_3$, which slightly increases the finesse without losing too much of the coupling.

\begin{figure}[ht]
\centering
   \includegraphics[width=0.8\linewidth]{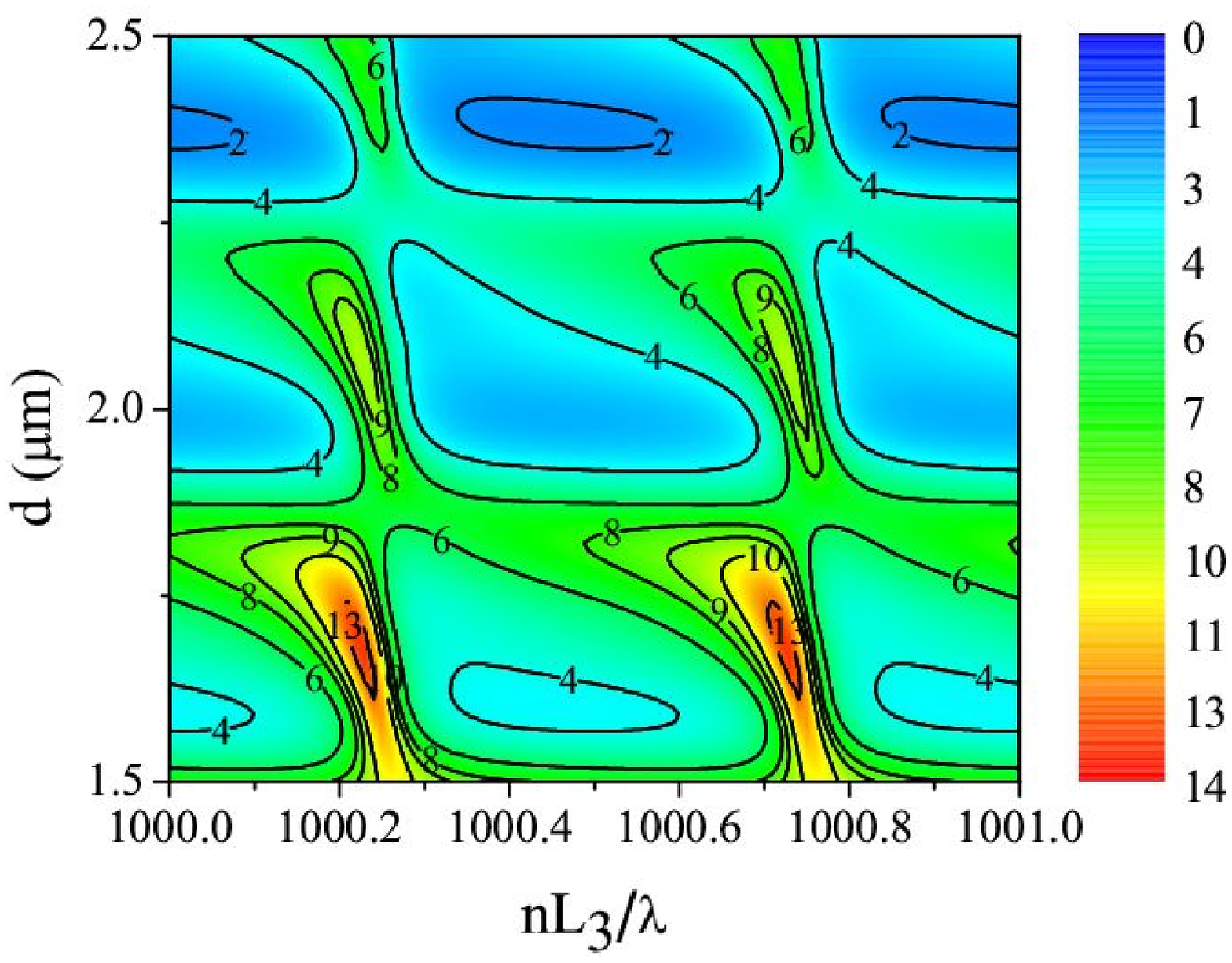}\\
  \caption{Cooperativity of a Rb atom in the coupled cavity system as a function of $d$ and $L_3$. The waveguides are assumed to be lossless.}
  \label{fig-Cfinal}
\end{figure}

We also performed simulations for a lossless waveguide, that is, in the situation where the only loss is diffraction in the gap ($\mathcal{F}_{d} \ll\mathcal{F}_{p}$). In this case the situation is completely different (see Fig. \ref{fig-Cfinal}). Optimizing the cooperativity corresponds to maximizing the finesse rather than to maximizing the coupling. To illustrate this point, consider a gap width $d = 1.95~\mu$m. For $nL_{3} = 1000.25 \lambda$, $g^{2}$ is $n^{2}\sim$10 times smaller than for $nL_{3} =$ 1000$\lambda$. However, the finesse is $\sim$29 times higher for $nL_{3} =$ 1000.25$\lambda$ and this choice clearly results in a higher $\mathcal{C}$. By setting $d =$ 2.05 $\mu$m and $nL_{3} =$ 1000.23$\lambda$, one can further increase the finesse and the coupling, reaching a cooperativity $\mathcal{C}_{max} \approx$ 9.4.

It is also interesting to notice that the maximum of $\mathcal{C}$ is very narrow with respect to $L_3$, but is much wider with respect to $d$. In the vicinity of $d\approx$ 1.95 $\mu$m, the $\mathcal{C}> 9$ area is 180 nm wide along the $d$ axis. This is important experimentally, because unlike the waveguide parts of the cavity ($L_1$ and $L_3$), whose refractive index can be changed by changing the temperature or by injecting a current, the gap's optical length cannot be fine-tuned once the cavity has been fabricated. These calculations show that if the gap is fabricated within 90 nm of the optimal width, it will always be possible to reach $\mathcal{C} > 95 ~\%\mathcal{C}_{max}$ by slightly tuning the index of refraction of $L_3$.

Another important parameter characterizing the atom-cavity system is the saturation photon number \cite{Enk}, the photon occupation number in the cavity necessary to saturate the atomic transition, 
$N_{\mathrm{sat}}=({\gamma / g})^2$ \cite{Kimble}. The calculations in Sec.~\ref{section-field} show that typically the atom field coupling is at least 10 times the natural linewidth of the Rb transition ($\gamma = 2 \pi \times 3\, \mathrm{MHz}$). Thus the saturation photon number can vary between 0.01 and 0.001 as the gap width and the total cavity length are varied.

The saturation photon number is useful to characterize the cavity transmission in the presence of an atom.
If we adjust the cavity parameters to give maximum cooperativity, the coupling is minimal ($g\sim 2 \pi \times 30 \, \mathrm{MHz}$) while the finesse is about 1500.  
In this situation we have $\kappa \sim 2 \pi \times 34 \, \mathrm{MHz}$ 
close to the value of $g$, but much larger than the atomic damping rate.
Without an atom the cavity, the transmission spectrum close to resonance show a peak with a width of order $\kappa$. 
Over a broader spectrum of frequencies, the cavity exhibits a complex structure characteristic of a multimirror Fabry-P\'erot interferometer as analyzed in Ref.~\cite{Stadt}.
The insertion of an atom into the cavity causes a dip in the in the absorption spectrum when the light frequency is close to the atomic resonance \cite{Poldy}. 
The width of this dip is of order $\gamma$. 
Reference~\cite{Poldy} discusses the optimization of the signal to noise in such a situation and points out that the optimum signal to noise ratio is achieved when the photon number is equal the saturation photon number. 
Direct observation of the vacuum Rabi splitting is not possible because of the large 
cavity damping.

\section{Summary}

We have analyzed the cooperativity of a cavity consisting of an optical waveguide in which a gap is opened to let atoms interact with the cavity field. Because of the high index of refraction of the guided parts of the cavity, the Fresnel reflections at the interfaces are not negligible and the gap behaves like a low finesse Fabry-Perot inside the high finesse cavity. We have separately calculated the losses induced by the diffraction of the light in the gap, and the value of the atom-light coupling. Choosing the cavity parameters such that all sub-cavities are resonant with the electromagnetic field maximizes the atom field coupling, but this choice also maximizes the diffraction losses introduced by the gap. This choice is optimal as long as the losses are dominated by other losses, such as absorption or diffusion in the waveguide. If the non-diffractive losses are small enough, a higher cooperativity is obtained by choosing the gap size and cavity length to decrease the coupling strength and increase the finesse. Our simulation predicts that a finesse on the order of 1500 and cooperativity $\mathcal{C} \sim$ 9 should be possible. We also showed that the value of the cavity parameters, especially the gap width, are tolerant to small deviations. The possibility of tuning the waveguide lengths means that close to optimum performance can be obtained despite small errors in the gap width. 
The achievable cooperativity can attain values which should be of interest for the detection of single atoms on chips. 
Although the fabrication of a gap remains a great technical challenge and has not yet been achieved, the fully integrated nature of the device we have analyzed means that we can envision a system of several such structures fabricated in parallel on the same substrate. This would permit a structure similar to that in reference \cite{Kohnen} with the additional feature of cavity enhanced atom sensitivity.

\section*{Acknowledgments}

We thank Philippe Lalanne for helpful discussions and for providing us with propagation simulation software. We acknowledge financial support from the Institut Francilien pour la Recherche en Atomes Froids, from the CNANO program of the Ile de France, and from the ANR program ANR-09-NANO-039.

\end{document}